\documentclass[12pt]{article}
\usepackage{amsfonts}
\usepackage{amsmath}
\usepackage{pifont}
\usepackage{natbib}
\usepackage{color}

\usepackage{url}

\title{Probability, Statistics and Planet Earth, \\ I: Geotemporal covariances.}
\author{N. H. Bingham and Tasmin L. Symons}

\begin{document}
\def\R{\mathbb{R}}
\def\C{\mathbb{C}}
\def\Z{\mathbb{Z}}
\def\N{\mathbb{N}}
\def\Q{\mathbb{Q}}
\def\D{\mathbb{D}}
\def\Sp{{\mathbb{S}}}
\def\T{\mathbb{T}}
\def\hb{\hfil \break}
\def\ni{\noindent}
\def\i{\indent}
\def\a{\alpha}
\def\b{\beta}
\def\e{\epsilon}
\def\d{\delta}
\def\De{\Delta}
\def\g{\gamma}
\def\qq{\qquad}
\def\L{\Lambda}
\def\E{\cal E}
\def\G{\Gamma}
\def\F{\cal F}
\def\K{\cal K}
\def\A{\cal A}
\def\B{\cal B}
\def\M{\cal M}
\def\P{\cal P}
\def\Om{\Omega}
\def\om{\omega}
\def\s{\sigma}
\def\th{\theta}
\def\Th{\Theta}
\def\z{\zeta}
\def\p{\phi}
\def\m{\mu}
\def\n{\nu}
\def\Si{\Sigma}
\def\q{\quad}
\def\qq{\qquad}
\def\half{\frac{1}{2}}
\def\hb{\hfil \break}
\def\half{\frac{1}{2}}
\def\pa{\partial}
\def\r{\rho}
\def\Spd{{{\Sp}}^d}

\maketitle

\begin{abstract}
The study of covariances (or positive definite functions) on the sphere (the Earth, in our motivation) goes back to Bochner and Schoenberg (1940--42) and to the first author (1969, 1973), among others. Extending to the geotemporal case (sphere cross line, for position and time) was for a long time an obstacle to geostatistical modelling. The characterisation question here was raised by the authors and Mijatovi\'c in 2016, and answered by Berg and Porcu in 2017. Extensions to multiple products (of spheres and lines) follows similarly (Guella, Menegatto and Peron, 2016). We survey results of this type, and related applications e.g. in numerical weather prediction.
\end{abstract}

\ni {\bf MSC2010 Classification}: 60-02; 60G60; 62H11; 60G15.\\

\ni { \bf 1. Introduction}  

It has been known for millennia that the Earth is spherical, and since then our natural interest in predicting weather and climate phenomena has driven the study of the interaction between spherical geometry and randomness, surveyed here and in the sequel. 

When one considers stochastic processes parametrised by the Earth (plus time) the first question is naturally one of existence.  For the prototypical case of Brownian motion on $\Sp^d \times \R$, see the recent survey [BinMS].

The application of probability and statistics to the Earth naturally begins with means and (co-)variances, before progressing to distributions, with Gaussianity as a first approximation.  The study of means, and their changes over time, is obviously of prime importance because of climate change.  We defer this and many other matters to Part II of this survey, and address covariances here. 

In the case of Gaussian processes, characterised completely by their mean and covariance, the question is one of positive definiteness on spheres. In order to bring in time, we need to work in a geotemporal context -- on a product space, `sphere cross (half-)line', where the spherical factor is position on the Earth and the (half-)line factor is time.  Until recently, the study of positive definite functions was focussed on spheres and other two-point homogeneous Riemannian manifolds, but this is too restrictive -- the interaction across space and time is of prime importance for the study of the Earth's atmosphere and oceans, from the short term (micro: weather) to the medium term (meso: prediction of next year's crops etc.) to the long term (macro: climate, and {\it climate change}).

Our treatment hinges on three results.  The first, from 1940-42, is classical; the second and third, from 2016 and 2017, are much more recent. \\

\ni {\bf 2. The Bochner-Schoenberg theorem}

The study of positive definite functions on general metric spaces dates back to Schoenberg in 1938 [Sch1], and to spheres $\Sp^d$ in particular to Bochner and Schoenberg 1940-42 ([BocS] in 1940; [Sch3] in 1940, [Sch4] in 1942; Bochner [Boc2] in 1941, [Boc3] in 1942).  Later extensions to probability theory were made by Kennedy [Ken], Lamperti [Lam], Gangolli [Gan] and Bingham [Bin1, 2].  The importance of this classical area has been widely recognised by the geostatistical community recently, notably by Gneiting [Gne1, 2, 3].

It is mathematically almost immaterial whether one chooses to work with covariances or with correlations.  The first is more convenient for applications, our motivation here, as they retain the natural units of the data; the latter are scalars, and scale-free.  They differ here merely by a scale-factor, which we write below as $c \in (0,\infty)$. Another possibility is the {\sl incremental variance} (or variogram), which must be of {\sl negative type}.  This is preferable for some purposes; see [BinMS] for background and references. 

We restrict attention to the {\it isotropic} case, where the positive definite function $f$ is of the form 
$$
f({\bf x},{\bf y}) = g(d({\bf x},{\bf y})) \qquad ({\bf x},{\bf y} \in {\Sp}^d),
$$ 
where $d$ denotes geodesic distance on the sphere; here $g$ is called the {\it isotropic part} of $f$.  For convenience, we adopt the usual normalisation of taking the sphere to have radius 1; thus $f$ is defined on $\Sp$, $g$ on $[-1,1]$, and $x = d({\bf x}, {\bf y}) \in [-1,1]$.  It turns out that the general case can be simply built up from a countable set of special cases.  The relevant $f$ here are {\it spherical harmonics}, or (zonal) {\it spherical functions} (\S 8.1, [BinMS], [Hel1,2]).  The relevant $g$ here are orthogonal polynomials (for background on which see e.g. [Sze]), the {\it Gegenbauer} (or {\it ultraspherical}) {\it polynomials}, of index
$$
\lambda = \half (d-1),
$$
on ${\Sp}^d \subset {\R}^{d+1}$ as above.  For the 2-sphere in 3-space (i.e. the Earth), these reduce ($\lambda = \half$) to the familiar {\it Legendre polynomials}, $P_n(x)$.  For the circle (1-sphere) in the plane, they reduce to the {\it Tchebycheff polynomials} 
$$
T_n(x) = \cos (n \ {\cos}^{-1} x) \qquad (x \in [-1,1]).
$$
All these are orthogonal polynomials with orthogonality interval $[-1,1]$ [Sze]. \\  

\ni {\bf  Bochner-Schoenberg Theorem}.  For ${\Sp}^d$, the general isotropic positive-definite function, equivalently, the general isotropic covariance, is given (to within a scale factor) by {\it a convex combination of Gegenbauer polynomials} $P_n^{\lambda}(x)$:
$$
c \sum_{n=0}^{\infty} a_n P_n^{\lambda}(x), \qquad a_n \geq 0, \quad \sum_{n=0}^{\infty} a_n = 1,                     \eqno(BS)
$$
with Legendre polynomials when $d = 2$. \\

\i In probabilistic language, this says that the general case is a {\it mixture} of spherical harmonics.  In particular, taking the mixing law to concentrate at one point: the Gegenbauer (Legendre) polynomials are themselves positive definite [Sch4, 177].  The gist of the Bochner-Schoenberg theorem is thus that {\it the Fourier-Gegenbauer coefficients} $a_n$ occurring in this mixture {\it are non-negative}.     

Askey and Bingham [AskB, \S 2] extended this result to Gaussian processes on the other compact symmetric spaces of rank 1 (two-point homogenous Riemannian manifolds).  \\

\ni {\bf 3. The Berg-Porcu theorem} \\

Recall some basic closure properties of the class ${\cal P} (M)$ of positive definite functions on a symmetric space $M$ (or semigroup, as in Berg et al. [BerCR]): ${\cal P}(M)$ is closed under positive scaling and convex combinations (both touched on in the above).  Further, ${\cal P}(M)$ is closed under pointwise products, which corresponds to convolution (of probability measures), and to tensor products (of representations); see e.g. [AskB, 5.2].  Similarly for product spaces (as here): by the Schur product theorem (see e.g. [HorJ, 458]), if $f_1 \in {\cal P}(M_1)$, $f_2 \in {\cal P}(M_2)$, then $f_1 f_2 \in {\cal P}(M_1 \times M_2)$.

We turn now from sphere $\Sp^d$ to sphere cross line $\Sp^d \times \R$.  As before, we confine attention to the isotropic case for the space variable on the sphere, and similarly, we restrict to the stationary case for the time-variable on the line.  The Bochner-Schoenberg theorem gives us a full knowledge of ${\cal P}(\Sp^d)$, and we know ${\cal P}(\R)$ (or ${\cal P}({\R}_+)$) from Bochner's theorem [Boc1] of 1933 for characteristic functions on the line.  By taking products, we obtain elements of ${\cal P}(\Sp^d \times \R)$.  These are covariances on sphere cross line, {\sl but} where the spatial and temporal aspects are independent.  These -- called {\it separable} space-time covariances by Gneiting [Gne2] -- are obviously unrealistic: for space-time modelling, one needs {\it non-separable} covariances, not of product form.  By the above, we can write down a large class of such functions by forming {\it convex combinations of separable} ones -- the class ${\cal P}(\Sp^d \times \R)$ being closed under convex combinations.  In view of the Bochner-Schoenberg theorem, the natural conjecture here is that this gives the general case; in any case, it gives a rich enough class to be amply flexible enough for modelling purposes (at least in the first instance).  With this in mind, the question was raised by Mijatovi\'c and the authors [BinMS, 4.4] in 2016 of extending the results of Askey and Bingham [AskB] to sphere cross line.  This has been done by Berg and Porcu [BerP], 2017.  Their result confirms the conjecture above: \\

\ni {\bf Berg-Porcu Theorem}.  The class of isotropic stationary sphere-cross-line covariances coincides with the class of convex combinations of separable ones:
$$
c \sum_{n=0}^{\infty} a_n P_n^{\lambda}(x) {\phi}_n(t), \qquad a_n \geq 0, \quad \sum_{n=0}^{\infty} a_n = 1. \eqno(BP)
$$
with the ${\phi}_n$  characteristic functions on the 
line.\\

\i Comparing this with the Bochner-Schoenberg theorem, one can readily conjecture what will happen with multiple products, of a number of spheres and of (half-)lines, representing time(s), height(s), etc.  This is the content of the {\it Guella-Menegatto-Peron theorem}, to which we now turn. \\

\ni {\bf 4.  Multiple products: the Guella-Menegatto-Peron theorem} \\

\i An enormous quantity of geostatistical data is collected for use in weather forecasting (\S 6 below) -- current operational weather forecast models use $\mathcal{O}(10^7)$ observations of rainfall, temperature, pressure, snow depth (when relevant), wind direction, speed and strength of gusts, visibility, cloud (type, amount, height), sunshine, etc, which are then combined with a dynamical model based on atmospheric fluid dynamics [Val].  The multivariate nature of this data motivates the extension of the work above to products of spheres.  This was done for two spheres in 2016 by Guella, Menegatto and Peron [GueMP1].  Their result extends, by their methods, to multiple products of spheres, and also to multiple products of lines, by the methods above.  We summarise all this as follows:  \\

\ni {\bf Guella-Menegatto-Peron Theorem}.  For a multiple product of $N$ spheres and $M$ (half-)lines, the general positive definite function isotropic in the space variables and stationary in the time variables is a multiple of a convex combination of separable ones,
\begin{align*}\tag{GMP}
c \sum_{n_1, \ldots, n_{M+N} =0}^{\infty} a_{n_1} 
P_{n1}^{{\lambda}_1}(x_1) \cdots P_{n_N}^{{\lambda}_r}(x_r) 
{\phi}_{n_{N+1}}(t_1) \cdots {\phi}_{n_{N+M}}(t_s),
\end{align*}
where $a_n \geq 0, \quad \sum_{n=0}^{\infty} a_n = 1$, as in the above. 

\i The authors of [GueMP1] remark (p.672-3) that they do not have any practical applications to offer.  In fact, their result is very timely and very rich in practical applications, to which we turn in \S 6. \\
\i The 1-sphere is relevant here for capturing {\it seasonal} effects, wind direction etc. \\
\i A related extension of the Berg-Porcu theorem to the multivariate case, motivated as here by Planet Earth, has recently been given by Alegria, Porcu, Furrer and Mateu [AlePFM]. \\

\ni {\bf 5.  Implementation: three kinds of truncation} \\

As with any result involving an infinite operation (summation, integration etc.), $(BP)$ calls for trunction in general before implementation can take place. \\

\ni {\it Index truncation}. \\
\i In Fourier analysis, the trigonometric functions (or complex exponentials) occur as eigenfunctions with eigenvalue $-n^2$.  This enables most Fourier series to be truncated for numerical use after comparatively few terms; for background, see e.g. the journal {\sl Applied and Computational Harmonic Analysis}.  With the ultraspherical polynomials, the corresponding eigenvalues are $-n(n + 2 \lambda)$ [Sze, (4.7.4)], which increase as fast.  So, truncation of the Fourier-Gegenbauer series after a suitable number of terms is as straightforward as in the Fourier case. \\
\i As one may expect the dominance of the early harmonics to be reflected in decreasing size of the coefficients $a_n$, one may use the technique of {\it monotone regression} to build this in to the model.  For details on this, see [BarBBB], [RobWD], and for the power of this technique in action, see e.g. [BinP]. \\  

\ni {\it Spatial truncation}. \\
\i The argument $x$ in $(BS)$, $(BP), (GMP)$ is in $[-1,1]$, and is
$$
x = \cos (d({\bf x}_1, {\bf x}_2)),
$$
with $d$ geodesic distance and ${\bf x}_i$ points on the surface of the Earth, normalised as above to have radius 1 (cf. e.g. [AskB, 133]).  Using the further normalisation
$$
W_n(x) := P_n^{\lambda}(x)/P_n(1),
$$
the spherical functions $W_n$ are uniformly bounded [AskB, 130], and
$$
|W_n(x)| \leq W_n(1) = 1
$$
for all $n$ and $x$.  Thus the spatial terms $P_n^{\lambda}(x)$ in $(BS)$, $(BP)$, $(GMP)$ are largest (in modulus) when their arguments, the two spatial points ${\bf x_i}$, are close.  So these give the most significant terms, and it makes sense to restrict also to nearby pairs of points when truncating (how close will depend on context; see below).\\

\ni {\it Temporal truncation}. \\
\i Weather (as distinct from seasonal or climatic aspects) is notoriously short-term: one might say that the system generating the Earth's weather has short-term memory.  It thus makes sense also to truncate the expansions to drop terms where the time-separation between the two space-time points is too long. \\
\i One popular method in geostatistical modelling is to model the time decay by a member of the Cauchy scale family with density and characteristic function
$$
f(x) = \frac{c}{\pi (1 + c^2 x^2)}, \qquad \phi(t) = \exp\left\lbrace {-|t|/c} \right\rbrace.
$$
Thus temporal decay like an inverse square seems appropriate here -- one may expect both short- and long-range dependence. \\

\ni {\it Combination}. \\
\i For practical use (below), all three forms of truncation -- index, spatial and temporal -- will be needed; which order they should be done in, and how, will depend on the type of data being modelled and the purpose of the study.  But $(BP)$ and $(GMP)$ give one all the flexibility one could ask for here. \\
     
\ni {\bf 6.  Applications: Kriging and Numerical Weather Prediction (NWP)} \\

These strikingly simple and very recent results, the Berg-Porcu  and Guella-Menegatto-Peron theorems, are of profound importance in several major geostatistical areas.\\

\ni {\it Kriging.}
 
Kriging (named after the mining engineer D. G. Krige) is the statistical technique for probabilistically interpolating a (space/ space-time) field from measured observations. An estimated point $Z(x)$ is formed of a weighted average (convex combination) of the observations $(Z(x_i))_{i=1}^n$, with closer observations typically given more weight. Clearly, then, the success of the interpolation procedure is dependent on the correct specification of the spatial structure of the field (in practice, knowledge of the incremental variance is necessary).  For details and references, see Cressie [Cre], Stein [Ste], and for kriging on the sphere, Young [You]. \\

\ni{\it Variational Data Assimilation}

Central to the process of constructing weather forecasts is {\sl data assimilation} -- the art and science of combining an imperfect model with inaccurate observations; see the books by Kalnay [Kal], Evensen [Eve2] and Reich and Cotter [ReiC] for overviews of the possible techniques here. To initialise the numerical model which generates a weather forecast, one needs the best possible approximation of the current state of the atmosphere (or ocean, or both) -- in the parlance of NWP this is known as the {\sl analysis} $\mathbf{x}_a$. Given a prior (or {\sl background}) $\mathbf{x}_b$, data assimilation seeks $\mathbf{x}_a | \mathbf{y}_o$, where the $\mathbf{y}_o$ is a noisy set of observations. Bayes's theorem gives this a probabilistic flavour:
$$
\text{p}(\mathbf{x} | \mathbf{y}_o) = \frac{ \text{p}(\mathbf{y}_o | {\bf x}) \text{p}_B({\bf x})}{\text{p}(\mathbf{y}_o)}.
$$
We seek to maximise $\text{p}({\bf x} | \mathbf{y}_o)$. Assuming $\mathbf{x}_b = \mathbf{x}_a + \epsilon_B$, $\epsilon_B \sim \text{N}(0, \bf{B})$ and $\mathbf{y}_o = H(\mathbf{x}) + \epsilon_O$, $\epsilon_O \sim \text{N}(0, \bf{R})$ this amounts to minimising
$$
J(\mathbf{x}) 
= (\mathbf{x} - \mathbf{x}_b)^T {\bf B}^{-1} (\mathbf{x} - \mathbf{x}_b) + (\mathbf{y}_o - H(\mathbf{x}))^T {\bf R}^{-1}(\mathbf{y}_o - H(\mathbf{x})).                                                                         \eqno(J)
$$
This is known as the {\sl cost function} of 3D-Var data assimilation -- see the textbooks by Daley [Dal] and more recently Kalnay [Kal] for details here. Improvements in this area have formed one strand of the `quiet revolution' in numerical weather prediction over the past decade [BauTB]. In particular, the uptake of four-dimensional assimilation (4D-Var) in operational settings allows for increased use of satellite data -- here, observations are allowed to arrive throughout a time interval, rather than (as in 3D-Var) being assumed coincidental with the desired time of analysis ${\bf x}_a$. This is especially important for forecasting in the Southern Hemisphere, with its smaller landmass (and therefore fewer traditional weather observation posts).  There are several complications to discuss, we focus on three:\\

\ni {\sl Size}.

With millions of observations ($\mathcal{O}(10^7)$/day) and billions of model variables ($\mathcal{O}(10^9)$), both $\mathbf{B}$ and $\mathbf{R}$ are huge. Indeed, the covariance matrix $\mathbf{B}$ has $\mathcal{O}(10^{18})$ entries, and is simply too big to be stored on a computer. In practice, (an approximation to) $\mathbf{B}$ is decomposed into more manageable matrices.\\

\ni{\sl Observation error correlations}. 

\i Until very recently [SteDN] the errors in the observations have been assumed to be independent -- that is, the observation error covariance matrix $\mathbf{R}$ has been assumed to be diagonal. To satisfy this assumption the (expensive to obtain) data may be thinned or `superobbed' (a kind of averaging). With the uptake of higher-resolution measurement instruments this situation has evolved from undesirable to untenable, and efforts are being made to understand and implement non-diagonal $\mathbf{R}$ matrices, with notable contributions by Desroziers et al. [DesBCP], and a school at the University of Reading [SteDN, WalDN, WalLDN], the latter of which is applied to the Ensemble Kalman Filter (see below).

Since the observations errors are Gaussian processes parametrised by $\Sp^2$, or $\Sp^2 \times \R$, the theorems above give a complete characterisation of the possible forms of the covariance structure of these errors.\\

\ni{\sl Nonlinearity}.

The operator $H$ above, which maps model space into observation space, is assumed to be linear. This is an unrealistic assumption in the context of weather and climate. A possible solution is the Ensemble Kalman Filter (see below), but this still requires the prior to be Gaussian. To move forward more inventive approaches are needed -- some ideas and developments in this direction are discussed by van Leeuwen et al. in [vanLCR, Part 1], including using variational data assimilation to arrive at good proposal densities, and new types of approximate particle filters which borrow from the ensemble Kalman filter.\\

\ni {\it Ensemble Kalman filter}\\
\i A recent development in numerical weather prediction is the idea of a probabilistic, rather than deterministic, forecast. ``Weather forecasts today involve an ensemble of numerical weather predictions, providing an inherently probabilistic assessment'' [BauTB, p. 48].\\
\i The ensemble Kalman filter (EnKF), introduced by Evensen in 1994, is a combination of the Kalman filter of control engineering (for background on which see e.g. Whittle [Whi]) with Monte Carlo methods: an ensemble of simulations (particles) are run, enabling one to approximate a true covariance matrix by a sample covariance matrix obtained from these.  For a monograph treatment see e.g. Evensen [Eve2], and for an earlier review paper, [Eve1]; it involves a combination of the Kalman filtering ideas above, Bayesian updating, traditional variational techniques (Euler-Lagrange equations) [Eve2, Ch. 9], and variational data assimilation (3D-Var for space, 4D-Var for space-time).  The method proved its practical worth early in work by Evensen and van Leeuwen [EvevL]  on the Agulhas Current (off the SE coast of S. Africa). The ensemble approach produces a spread of results, or scenarios, which capture the uncertainty inherent in weather forecasting and allow a probabilistic interpretation by forecasters.\\
\i The dominant store of energy in the environment is in the oceans; this is the energy source that drives the weather.  EnKF has been widely used for study of the oceans; see e.g. [Eve2, Ch. 16].  It is already finding use in atmospheric studies; its use in coupled ocean-atmosphere interaction is crucially important, but yet to be fully developed. \\ 

\i For an application of data assimilation to three-dimensional hydrodynamics, see Wolf et al. [WolSBW]. \\

\ni {\bf 7. Chordal and geodesic distance.} \\

A function $f: \R^{d+1} \times \R^{d+1} \to \R$ is positive definite [BerCR] if 
$$
\sum_{i, j = 1}^n c_i c_j f(x_i, x_j) \geq 0
$$
for all finite collections $x_1, \ldots, x_n \in \R^{d+1}$ and real constants $c_1, \ldots, c_n \in \R $.  These $f$ are the Fourier(-Stieltjes) transforms of (positive, finite) measures, by  Bochner's theorem.  If, further,$f(x,y)$ is a function purely of the (Euclidean) distance $ t = ||x - y||$ between its two arguments, rather than the orientation of $x - y$, we say $f$ is {\sl isotropic}. Such functions arise by radialization of the Fourier transform; they have an integral representation due to Schoenberg [Sch2]:
$$
f(t) = \int_0^\infty \Gamma \left(\frac{d-1}{2}\right) \left({\frac{2}{rt}}\right)^{\frac{d-1}{2}} J_{(d-1)/2)} (rt) \, \text{d}\mu (r),
$$
where $\mu$ is a probability measure on $\R^+$ and $J_{(d-1)/2}$ is a Bessel function of the first kind (cf. Bochner and Chandrasekharan [BocC, II.7]).  These modified Bessel functions arise whenever one radializes a characteristic function, as in Kingman's random walks with spherical symmetry [Kin]. 


A natural question here is whether one can restrict these functions from $\R^{d+1}$ to the sphere $\Sp^d$ whilst retaining positive definiteness. A recipe due to Yadrenko [Yad] allows for the construction of a positive definite function $\hat{f}$ on $\Sp^d$ from a positive definite function $f$ on $\R^{d+1}$ via the transformation 
$$
\hat{f}(\theta) = f \left( 2 \sin \frac{\theta}{2} \right), \qquad \theta \in [0, \pi]. 
$$
Although this construction is convenient, one predictably pays a price for ignoring the geometry of the sphere.  Gneiting [Gne3] highlighted this approach's severe restrictions -- there is a lower bound of $\inf_{t>0} 1/t \sin t \approx -0.21$ on $\hat{f}$ when $d=2$ (the geostatistical case). Moreover, Gneiting argues that, since  $\sin \theta \approx \theta$ when $\theta$ is small, this construction is doomed to conflict with spherical geometry. It is preferable, therefore, to construct our positive definite functions on the sphere itself, using {\sl geodesic} (or great circle) distance in preference to Euclidean distance, as we have done above. \\

\ni {\bf 8.  Complements} \\

\ni {\it 8.1.  Spherical harmonics, spherical functions, Gegenbauer polynomials}. \\
\i Spherical harmonics play the role in harmonic analysis on the sphere that trigonometric functions (or complex exponentials) play on the line.  They have been extensively studied, in classics such as MacRobert [MacR], Hobson [Hob] and more recently in e.g. Muller [Mul], Andrews, Askey and Roy [AndAR, Ch. 9]. \\
\i Spherical functions (the term `zonal' used to be added, reflecting the isotropy restriction we used) arise similarly, as the name suggests.  But they can be defined much more generally, in the context of symmetric spaces; see e.g. [Hel1,2], [BinMS] for background and references.  In the compact case, as with spheres, these form a countable set.  These reduce, as here, to the Gegenbauer (or for us, Legendre) polynomials.  In turn, the addition formula for the Gegenbauer polynomials reflects (and is most easily proved via) the action of the group $SO(d)$ on the sphere ${\Sp}^d = SO(d+1)/SO(d)$ as a coset space (symmetric space, as above); see [BinMS].  \\
\i The reason that the mathematics here is dominated by orthogonal polynomials rests directly on the geometry of the sphere: it arises as a quotient of compact Lie groups; harmonic analysis on these rests on the Peter-Weyl theorem, for which one has the Schur orthogonality relations; see e.g. [App, \S 2.2.1]. \\
   
\ni {\it 8.2.  The Funk-Hecke formula and Fourier-Gegenbauer coefficients}. \\
\i The Funk-Hecke formula ([Mul], [AndAR Ch. 9]) relates (in the notation of \S 2) the integral in $g$ over $[-1,1]$ in the Fourier-Gegenbauer coefficients to an integral in $f$ over the sphere ${\Sp}^d$.  It is the positivity of these that is the crux here. \\ 
   
\ni {\it 8.3.  Strict positive definiteness}. \\
\i {\it Strict} positive definiteness is defined in the usual way: the relevant (non-negative) sums are {\it positive} except when all the coefficients vanish.  Chen, Menegatto and Sun [CheMS] and Barbosa and Menegatto [BarM] characterise {\it strict} positive definiteness here: infinitely many coefficients $a_n$ need to be positive (for both even and odd $n$ in the case of spheres, a complication reflecting antipodal points).  Similar studies have been carried out by Guella, Menegatto and Peron ([GueM], [GueMP2,3]).  Interesting though these results are, they are of limited practical importance: the strictness criterion (infinitely many positive coefficients) does not survive the truncation needed for practical implementation (\S 6).    \\

\ni{\sl 8.4. The nugget effect}. \\
\i The nugget effect refers to the phenomenon of a non-zero incremental variance at zero separation, initially observed (and named after) the large amount of variability in observations of gold nuggets in very small geographic areas. The existence of the nugget -- or more precisely, whether it is an artefact of measurement error or whether it is inherent micro-scale variability in the process being observed -- is a subject of some scientific debate. See [Cla] for more details of this interesting effect, and [GneGG] for an example of the nugget effect being fitted in a spatio-temporal covariance model. \\

\ni{\sl 8.5. Lagrangian frameworks}.\\
\i In fluid dynamics one has a choice of one's frame of reference -- one may either sit on the river bank and watch the water pass (the {\sl Euclidean} framework), or one may follow the body of water in a boat (the {\sl Lagrangian} framework). In geostatistical cases, when symmetry is an inappropriate assumption, a Lagrangian covariance structure may be more suitable, or, as suggested by Gneiting [Gne2], a convex combination of a symmetric covariance and a Lagrangian one.

The basic idea of a Lagrangian covariance is to take a stationary spatial random field with covariance $C_s({\bf h})$ and follow it along as it moves with (random) velocity ${\bf V}$.  The covariance of this field then has the form
$$
C({\bf h}, t) = \mathbb{E}_{\bf V} \left( C_s({\bf h} - t{\bf V}) \right).
$$
Gneiting et al. [GneGG] discuss possible choices for $\mathbf{V}$, in particular noting that prevailing winds may be modelled by the simplest case $\mathbf{V} = \mathbf{v}$, a constant. They also raise the interesting possibility of a dynamic velocity $\mathbf{V}(t)$, which would yield nonstationary covariances. This framework has recently been extended to the $\Sp^d \times \R$ case by Alegria and Porcu [AleP1], who also discuss the {\sl dimple effect} (where the field is more strongly correlated when one steps forwards in time than than space) for transport models.\\

\ni{\sl 8.6. Seasonality}. \\
\i Taking $\R$ as the time domain of a spatio-temporal field is natural enough, but doing so restricts us to cases where there is no cyclical component to avoid violating the assumption of stationarity. 

An intriguing approach by Alegria and Porcu [AleP2] combines the idea of the Guella-Menegatto-Peron multiple product $\Sp^d \times \Sp^d$ (\S 4) with the work of Berg and Porcu on $\Sp^d \times \R$ (\S 3) to model spatio-temporal fields on $\Sp^d \times \Sp^1 \times \R$ -- with the temporal component decomposed into $\Sp^1 \times \R$ the cyclical component can be dealt with separately in a very natural way. The approach is complemented by an extension to the Lagrangian framework, as done in the $\Sp^d \times \R$ case by the same authors [AleP1].\\

\ni {\it 8.7.  Mixture models}. \\
\i All three of the Bochner-Schoenberg, Berg-Porcu and Guella-Menegatto-Peron theorems give the relevant classes of covariances as (multiples of) {\it mixtures} (convex combinations), of an $n$th harmonic -- $n$th degree Legendre (or Gegenbauer) factor, times other such factors -- separable covariances, in the BP and GMP cases.  For implementation, we must truncate the infinite mixture to a finite one.  Such (finite) {\it mixture models} have been extensively studied in statistics, and a great deal is known about them.  Our main source is the book of McLachlan and Peel [McLP]; see also Titterington, Smith and Makov [TitSM], Lindsay [Lind]. \\
\i For large data sets as here, computer implementation is essential.  The main tool here is EMMIX [McLP], a version for mixture models of the EM (expectation-maximization) algorithm of Dempster, Laird and Rubin [DemLR]; see also Meng and van Dyk [MenvD].  One adapts this here from mixtures of distributions to mixtures of covariance matrices by replacing variational distance by a suitable matrix norm. \\
\i We note that the relevant covariance matrices are of block diagonal form, reflecting the separability mentioned above.  Their inversion together is thus no harder than their inversion separately.  We are also spared having to model off-diagonal block submatrices representing, say, space-time interaction: this is done for us by the Legendre-polynomial (`spherical harmonic') structure in $(BS), (GMP)$. \\
\i In regression, one typically decides how many parameters one is prepared to use, and then tests the hypothesis that the last one used is statistically significant (Kolodzieczyk's theorem: see e.g. [BinF, Th. 6.5]).  In deciding how many misture components (harmonics) to retain here, one can proceed as there, or use a Bayesian approach, putting a prior on the number of components.  For various choices here, see [McLP, \S 4.7]. \\ 
\i Mixture ideas are used in the Euclidean case by Ma [Ma] and Porcu and Mateu [PorM]. \\  

\ni {\it 8.8. Large covariance matrices}. \\
\i In $(J)$ of \S 6, one needs to invert large sample-covariance matrices $B$, $R$. 
If not sparse enough, this is a challenging numerical task; however, there is a great deal of both relevant theory and practical experience.  For the latter, we refer to the sources cited in \S 6, and the current practices of data thinning and `superobbing' (averaging over groups of data, and inflating covariances) in NWP.  For the former, one needs the {\it condition number} (ratio of largest to smallest eigenvalues: see e.g. Wilkinson [Wil, \S 4.3], Golub and Van Loan [GolVL, Ch. 8]); in this context, see e.g. Bai and Lin [BaiL], Lin, Bai and Krishnaiah [LinBK].  Among a wealth of other sources, we mention here also Pourahmadi [Pou1,2], Ledoit and Wolf [LedW1,2], Won et al. [WonLKR], Bickel and Levina [BicL], El Karoui [ElK].  \\

\ni {\it 8.9.  Statistics on spheres and spherical regression}. \\
\i Statistics on the 1-sphere (circle) is treated in Fisher [Fis].  One motivation (Ch. 7) is wind direction.  Statistics on the sphere is treated in Watson [Wat] and Fisher, Lewis and Embleton [FisLE].  One needs a range of spherical analogues of the standard distributions in Euclidean space, including the Fisher, Fisher-Bingham (R. A. Fisher and C. Bingham) and Kent distributions. A classical motivation here was to remanent magnetism in rocks, seeking support for Wegener's theory of continental drift.  For an application of mixture models to joint sets (parallel fractures) in rocks, see Peel et al. [PeeWM].  For spherical regression, similarly applicable to plate tectonics, see Downs [Dow] and the references cited there. \\   

\ni {\it 8.10.  Schur polynomials}. \\
\i A recent algebraic approach to this area, also motivated by such things as the Berg-Porcu theorem and large covariance matrices, has been given by Belton, Guillot, Khare and Putinar [BelGKP].  It uses Schur polynomials, and properties of symmetric functions; for details, see [BelGKP] and the references cited there.\\

\ni {\bf Acknowledgement}. \\

\i The second author thanks the EPSRC Mathematics of Planet Earth CDT, based jointly in the Mathematics Department at Imperial College, London and the Meteorology Department at Reading University, for financial support during her doctoral studies.  The first author thanks the same sources for rekindling his mathematical interest in his own doctoral studies long ago. \\

\ni{\bf References} \\

\ni [AleP1]  A. Alegria and  E. Porcu. The dimple problem related to space-time modelling under a Lagrangian framework. \url{arXiv: 1611.09005v1} (2016).\\
\ni [AleP2]  A. Alegria and  E. Porcu. Space-time geostatistical models with both linear and seasonal structures in the temporal components. \url{arXiv: 1702.01400} (2017).\\
\ni [AlePFM] A. Alegria, E. Porcu, R. Furrer and J. Mateu. Covariance functions for multivariate Gaussian fields evolving temporally over Planet Earth, \url{arXiv:1701.00601v1} (2017). \\
\ni [AndAR] G. E. Andrews, R. Askey and R. Roy, {\sl Special functions}.  Encycl. Math. Appl. {\bf 71}, Cambridge University Press, 1999. \\
\ni [App] D. Applebaum, {\sl Probability on compact Lie groups}.  Springer, 2014. \\
\ni [AskB] R. A. Askey and N. H. Bingham, Gaussian processes on compact symmetric spaces.
{\sl Z. Wahrschein.  verw. Geb.} {\bf 37} (1976), 127-143. \\
\ni [BaiL] Z. D. Bai and Y. Q. Lin, Limit of the smallest eigenvalue of a large-dimensional sample covariance matrix.  {\sl Ann. Probab.} {\bf 21} (1993), 1275-1294. \\
\ni [BarM] V. S. Barbosa and V. A. Menegatto. Strictly positive definite kernels on two-point compact homogeneous spaces. {\sl Math. Inequal. Appl.} {\bf 19}(2) (2015), 743--756. \\
\ni [BarBBB] R. E. Barlow, D. J. Bartholomew, J. M. Bremner and H. D. Brunk, {\sl Statistical inference under order restrictions}.  Wiley, 1972. \\
\ni [BauTB] P. Bauer, A. Thorpe and G. Brunet. The quiet revolution of numerical weather prediction. {\sl Nature} {\bf 525} (2015), 47--55. \\
\ni [BelGKP] A. Belton, D. Guillot, A. Khare and M. Putinar, Schur polynomials and matrix positivity preservers.  \url{arXiv:1602.04777} (2016).\\
\ni [BerCR] C. Berg, J. P. R. Christensen and P. Ressel, {\sl Harmonic analysis on semigroups: Theory of positive definite and related functions}.  Springer, 1984. \\
\ni [BerP] C. Berg and E. Porcu, From Schoenberg coefficients to Schoenberg functions. {\sl Constructive Approximation} {\bf 45} (2017), 217 -- 241.\\
\ni [BicL] P. J. Bickel and E. Levina, Regularised estimation of large covariance matrices.  {\sl Ann. Statist.} {\bf 36} (2008), 188-227. \\
\ni [Bin1] N. H. Bingham, {\sl Limit theorems and semigroups in probability theory.} PhD thesis, University of Cambridge, 1969. \\
\ni [Bin2]  N. H. Bingham, Positive definite functions on spheres.  {\sl Proc. Cambridge Phil. Soc.} {\bf 73} (1973), 145-156. \\
\ni [BinF] N. H. Bingham and J. M. Fry, {\sl Regression: Linear models in statistics}.  Springer, 2010. \\
\ni [BinMS] N. H. Bingham, Alexsandar Mijatovi\'c and Tasmin L. Symons. Brownian manifolds, negative type and geo-temporal covariances. {\sl Communications in Stochastic Analysis (H. Heyer Festschrift)}, {\bf 10}(4) (2016), 421 -- 432. \\
\ni [BinP]  (with S. M. Pitts): Non-parametric inference for the
$M/G/\infty$ queue.  {\sl Ann. Inst. Statistical Mathematics} {\bf 51} (1999), 71-97. \\
\ni [Boc1] S. Bochner, Monotone Funktionen, Stieltjessche Integrale und harmonische Analyse.  {\sl Math. Annalen} {\bf 108} (1933), 378-410. \\
\ni [Boc2] S. Bochner, Hilbert distances and positive definite functions.  {\sl Ann. Math.} {\bf 42} (1941), 647-656.  \\
\ni [Boc3] S. Bochner, Review of [Sch4]. {\sl Math. Reviews} {\bf 3} (1942) p.232 
(MR0005922). \\
\ni [BocC] S. Bochner and K. Chandrasekharan, {\sl Fourier transforms}.  Annals of Mathematics Studies 19, Princeton University Press, 1949. \\
\ni [BocS] S. Bochner and I. J. Schoenberg, On positive definite functions on compact spaces.  {\sl Bull. Amer. Math. Soc.} {\bf 46} (1940), 881. \\
\ni [Cla] I. Clark. Statistics or geostatistics? Sampling error or nugget effect? {\sl Journal of the Southern African Institute of Mining and Metallurgy} {\bf 110}(6) (2010), 307--312.\\
\ni[CheMS] D. Chen, V. A. Menegatto and X. Sun. A necessary and sufficient condition for strictly positive definite functions on spheres. {\sl Proc. Amer. Math. Soc.} {\bf 131}(9) (2003), 2733--2740.\\
\ni [Cre] N. Cressie, {\sl Statistics for spatial data}, revised ed., Wiley, 1993. \\
\ni [Dal] R. Daley. {\sl Atmospheric Data Analysis}. Cambridge University Press, 1991.\\
\ni [DemLR] A. P. Dempster, N. M. Laird and D. B. Rubin, Maximum likelihood from incomplete data via the EM algorithm (with discussion).  {\sl J. Royal Statistical Society B} {\bf 39}, 1-38.  \\
\ni [DesBCP] G. Desroziers, L. Berre, B. Chapnick and P. Poli. Diagnosis  of  observation, background  and  analysis-error  statistics  in  observation  space. {\sl Q. J. R. Met. Soc.} {\bf 131} (2005), 3385 -- 3396.\\
\ni [Dow] T. Downs, Spherical regression.  {\sl Biometrika} {\bf 90} (2003), 655-668. \\
\ni [ElK] N. El Karoui, Spectral estimation for large-dimensional covariance matrices using random-matrix theory.  {\sl Ann. Statist.} {\bf 36} (2008), 186-197. \\ 
\ni [Eve1] G. Evensen, The ensemble Kalman filter: theoretical formulation and practical implementation.  {\sl Ocean Dynamics} {\bf 53} (2003), 343-367. \\
\ni [Eve2] G. Evensen, {\sl Data assimilation: The ensemble Kalman filter}.  Springer, 2006. \\
\ni [EvevL] G. Evensen and P. J. van Leeuwen, Assimilation of Geosat altimeter data for the Agulhas current using the ensemble Kalman filter with a quasi-geostrophic model.  {\sl Monthly Weather Review} {\bf 24} (1996), 85-96. \\
\ni [FinHI] B. Finkenst\"adt, L. Held and V. Isham, {\sl Statistics of spatio-temporal systems}.  Chapman \& Hall/CRC, 2007. \\
\ni [Fis] N. I. Fisher, {\sl Statistical analysis of circular data}.  Cambridge University Press, 1993. \\
\ni [FisLE] N. I. Fisher, T. Lewis and B. J. J. Embleton, {\sl Statistical analysis of spherical data}.  Cambridge University Press, 1987. \\  
\ni [Gan] R. Gangolli, Positive definite kernels on homogeneous spaces and certain stochastic processes related to L\'evy's Brownian motion of several parameters.  {\sl Ann. Inst. H. Poincar\'e B (NS)} {\bf 3} (1967), 121-226. \\
\ni [Gne1] T. Gneiting. Correlation functions for atmospheric data analysis. {\sl Q. J. R. Met. Soc.} {\bf 135} no. 559 (1999), 2449 -- 2464. \\
\ni [Gne2] T. Gneiting, Non-separable, stationary covariance functions for space-time data.  {\sl J. Amer. Stat. Assoc.} {\bf 87} (2002), 590-600. \\
\ni [Gne3] T. Gneiting, Strictly and non-strictly positive definite functions on spheres.  {\sl Bernoulli} {\bf 19} (2013), 1327 - 1349. \\
\ni [GneGG] T. Gneiting, M. G. Genton and P. Guttorp, Geostatistical space-time models, stationarity, separability and full symmetry.  In [FinHI], 151 - 175. \\
\ni [GolVL] G. H. Golub and C. F. Van Loan, {\sl Matrix computation}, 3rd ed.  The Johns Hopkins University Press, 1996. \\
\ni [GueM] J. C. Guella and V. A. Menegatto, Strictly positive definite kernels on a product of spheres.  {\sl J. Math. Anal. Appl.} {\bf 435} (2016), 286-301. \\
\ni [GueMP1] J. C. Guella, V. A. Menegatto and A. P. Peron, An extension of a theorem of Schoenberg to products of spheres.  {\sl Banach J. Math. Analysis} {\bf 10} (2016), 671-685. \\
\ni [GueMP2] J. C. Guella, V. A. Menegatto and A. P. Peron, Strictly positive definite kernels on a product of spheres II.  {\sl SIGMA (Symmetry, Integrability and Geometry, Methods and Applications)} {\bf 12} (2016), 15p. \\
\ni [GueMP3] J. C. Guella, V. A. Menegatto and A. P. Peron, 
Strictly positive definite kernels on a product of circles.  {\sl Positivity} {\bf 21} (2017), 329-342. \\
\ni [Hel1] S. Helgason, {\sl Differential geometry and symmetric spaces}.  Academic Press, 
1962. \\
\ni [Hel2] S. Helgason, {\sl Differential geometry, Lie groups and symmetric spaces}.  Academic Press, 1978. \\
\ni [Hob] E. W. Hobson, {\sl The theory of spherical and elllipsoidal harmonics}.  Cambridge University Press, 1931. \\
\ni [HorJ] R. A. Horn and C. R. Johnson, {\sl Matrix analysis}, 2nd ed., Cambridge University Press, 2013. \\
\ni [Kal] E. Kalnay. {\sl Atmospheric modelling, data assimilation and predictability}. Cambridge University Press, 2003. \\
\ni [Ken] M. Kennedy, A stochastic process associated with the ultraspherical polynomials. {\sl Proc. Roy. Irish Acad. Sect. A} {\bf 61} (1969), 61 -- 89. \\
\ni [Kin] J. F. C. Kingman, Random walks with spherical symmetry.  {\sl Acta Math.} {\bf 109} (1963), 11-53. \\ 
\ni [Lam] J. Lamperti, The arithmetic of certain semigroups of positive operators. {\sl Proc. Cambridge Philos. Soc.} {\bf 64} (1969), 161 -- 166.\\
\ni [LedW1] O. Ledoit and M. Wolf, A well-conditioned estimate for large-dimensional covariance matrices.  {\sl J. Multivariate Analysis} {\bf 88} (2004), 365-431. \\
\ni [LedW2] O. Ledoit and M. Wolf, Non-linear shrinkage estimation of large-dimensional covariance matrices.  {\sl Ann. Statist.} {\bf 40} (2012), 1024-1060. \\
\ni [LinBK] Y. Q. Lin, Z. D. Bai and P. R. Krishnaiah, On the limit of the largest eigenvalue of the large-dimensional sample covariance matrix.  {\sl Probab. Th. Rel. Fields} {\bf 78} (1988), 509-521. \\
\ni [Lind] B. G. Lindsay, {\sl Mixture models: Theory, geometry and applications}.  NSF-CBMS Reg. Conf. Ser. Prob. Stat. 5, IMS, 1995. \\
\ni [Ma] C. Ma, Spatio-temporal covariance functions generated by mixtures.  {\sl Math. Geology} {\bf 34} (2002), 651-684. \\
\ni [MacR] T. M. MacRobert, {\sl Spherical harmonics}.  Methuen, 1927. \\
\ni [McLP] G. McLachlan and D. Peel, {\sl Finite mixture models}.  Wiley, 2000. \\
\ni [MenvD] X. L. Meng and D, van Dyk, The EM algorithm -- an old folk song sung to a new fast tune (with discussion).  {\sl J. Royal Statistical Society B} {\bf 59} (1997), 511-567. \\ 
\ni [MonAF] P Monestiez, D. Allard and R. Froidevaux. {\sl geoENV III : Geostatistics for Environmental Applications.
Proceedings of the Third European Conference on Geostatistics for Environmental Applications, held in Avignon, France, November 22–24, 2000}. Springer, 2000. \\
\ni [Mul] C. Muller, {\sl Spherical harmonics}.  Lecture Notes in Math. {\bf 17} (1966), Springer. \\
\ni [PeeWM] D. Peel, W. J. Whiten and G. McLachlan, Fitting mixtures of Kent distributions to aid in joint set identification.  {\sl J. Amer. Statist. Assoc.} {\bf 96} (2001), 56-63. \\
\ni [PorM] E. Porcu and J. Mateu, Mixture-based modelling for space-time data.  {\sl Environmetrics} {\bf 18} (2007), 285-302. \\
\ni [Pou1] M. Pourahmadi, Covariance estimation: the GLM and regularization perspectives.  {\sl Statistical Science} {\bf 26} (2011), 369-387. \\
\ni [Pou2] M. Pourahmadi, {\sl High-dimensional covariance matrices}.  Wiley, 2013. \\  
\ni [ReiC] S. Reich and C. Cotter, {\sl Probabilistic forecasting and Bayesian data assimilation}, Cambridge University Press, 2015. \\
\ni [RobWD] T. Robertson, F. T. Wright and R. L. Dykstra, {\sl Order-restricted statistical inference}.  Wiley, 1988. \\ 
\ni [Sch1] I. J. Schoenberg, Metric spaces and positive definite functions.  {\sl Trans. Amer. Math. Soc.} {\bf 44} (1938), 522-536 (reprinted in {\sl I. J. Schoenberg, Selected Papers, Vol. 1} (ed. C. de Boor), Birkh\"auser, 1988, 100-114). \\
\ni [Sch2] I. J. Schoenberg, Metric spaces and completely monotone functions. {\sl Ann. Math.} {\bf 39} (4) (1938), 811 -- 841. \\
\ni [Sch3] I. J. Schoenberg, On positive definite functions on spheres. {\sl Bull. Amer. Math. Soc} {\bf 46} (1940), 888. \\ 
\ni [Sch4] I. J. Schoenberg, Positive definite functions on spheres.  {\sl Duke Math. J. } {\bf 9} (1942), 96-108 ({\sl Selected Papers 1}, 172-185). \\ 
\ni [Ste] M. L. Stein, {\sl Interpolation of spatial data: some theory for kriging}.  Springer, 2012. \\
\ni[SteDN] L. M. Stewart, S. L. Dance and N. K. Nichols, Correlated observation errors in data assimilation. {\sl Int. J. Num. Meth. Fluids} {\bf 56} no. 8 (2007), 1521 -- 1527. \\
\ni [Sze] G. Szeg\H{o}, {\sl Orthogonal polynomials}, Amer. Math. Soc. Colloquium Publications XXIII, AMS, 1939. \\
\ni [TitSM] D. M. Titterington, A. F. M. Smith and U. E. Makov, {\sl Statistical analysis of finite mixture distributions}.  Wiley, 1985. \\
\ni [Val] G. K. Vallis, {\sl Atmospheric and oceanic fluid dynamics: Fundamentals and large-scale circulation}.  Cambridge University Press, 2006. \\
\ni [vanLCR] P. J. van Leeuwen, Y. Cheng and S. Reich. {\sl Nonlinear Data Assimilation}. Springer, 2015.\\
\ni [WalDN] J. A. Waller, S. L. Dance and N. K. Nichols. Theoretical insight into diagnosing observation error correlations using observation-minus-background and observation-minus-analysis statistics. {\sl Q. J. R. Met. Soc.} {\bf 142} (2016), 418 -- 431.\\
\ni [WalDLN] J. A. Waller, S. L. Dance, A. S. Lawless and N. K. Nichols. Estimating correlated observation  error  statistics  using  an  ensemble  transform  Kalman  filter. {\sl Tellus A} {\bf 66} (2014).\\
\ni [Wat] G. S. Watson. {\sl Statistics on spheres}. Wiley, 1983.\\
\ni [Whi] P. Whittle, {\sl Optimal control: Basics and beyond}.  Wiley, 1996. \\
\ni [Wil] J. H. Wilkinson, {\sl The algebraic eigenvalue problem}.  Oxford Univerity Press, 1965. \\
\ni [WolSBW] T. Wolf, J. S\'en\'egal, L. Bertinas and H. Wackernagel, Application of data assimilation to three-dimensional hydrodynamics: The case of the Odra lagoon. In [MonAF], 157--168. \\
\ni [WonLKR] J.-H. Won, J. Lim, S.-J. Kim and B. Rajaratnam, Condition-number regularised covariance estimation.  {\sl J. Royal Statistical Society B} {\bf 75} (2013), 427-450. \\
\ni [Yad] M. \v{I}. Yadrenko, {\sl Spectral theory of random fields.} Translation Series in Mathematics and Engineering. New York: Optimisation Software, 1983.\\
\ni [You] D. S. Young, Random vectors and spatial analysis by geostatistics for geotechnical applications.  {\sl Mathematical Geology} {\bf 19} (1987), 467-479. \\

\ni N. H. Bingham, Mathematics Department, Imperial College, London SW7 2AZ; n.bingham@ic.ac.uk\\

\ni Tasmin L. Symons, Mathematics Department, Imperial College, London SW7 2AZ; tls111@ic.ac.uk

\end{document}